\documentclass[review]{elsarticle}

\makeatletter
\def\ps@pprintTitle{%
 \let\@oddhead\@empty
 \let\@evenhead\@empty
 \def\@oddfoot{\centerline{\thepage}}%
 \let\@evenfoot\@oddfoot}
\makeatother

\usepackage{graphicx}
\usepackage[cmex10]{amsmath}
\usepackage{amssymb}
\usepackage{amsfonts,amsthm}

\def\diag{\mathop{\mathrm{diag}}}

\journal{Digital Signal Processing}

\begin{document}

\begin{frontmatter}

\title{Olfactory Signal Processing\tnoteref{mytitlenote}}
\tnotetext[mytitlenote]{Portions of the material in this paper were first presented in \cite{VarshneyV2014a} and \cite{VarshneyV2014b}.}

\author{Kush~R.~Varshney}
\address{Mathematical Sciences and Analytics Department\\ IBM Thomas J.\ Watson Research Center\\ Yorktown Heights, NY 10598 USA}
\author{Lav~R.~Varshney}
\address{Department of Electrical and Computer Engineering\\ University of Illinois at Urbana-Champaign\\ Urbana, IL 61801 USA}

\begin{abstract}
Olfaction, the sense of smell, has received scant attention from a signal processing perspective in comparison to audition and vision.  In this paper, we develop a signal processing paradigm for olfactory signals based on new scientific discoveries including the psychophysics concept of \emph{olfactory white}.  We describe a framework for predicting the perception of odorant compounds from their physicochemical features and use the prediction as a foundation for several downstream processing tasks.  We detail formulations for odor cancellation and food steganography, and provide real-world empirical examples for the two tasks.  We also discuss adaptive filtering and other olfactory signal processing tasks at a high level.
\end{abstract}

\end{frontmatter}

\section{Introduction}
\label{sec:introduction}

Audition, vision, and olfaction are the three ways that people remotely sense stimuli; much signal processing research has dealt with audio and video signals, but study of olfactory signal processing has been neglected.  One reason is the difficulty in compactly specifying the fundamental inputs to the human perceptual system.  Whereas vibration and light signals interacting with the ears and eyes are compactly parameterized by amplitude, phase, and frequency, olfactory signals interacting with the nose manifest as collections of chemical compound molecules drawn from a very large set. Despite the possible input set having very large cardinality, recent evidence suggests that the space of olfactory perception is fairly low-dimensional \cite{MamloukM2004,KoulakovKER2011,CastroRC2013}.  The most basic dimension, akin to the DC component of a waveform, is \emph{pleasantness} \cite{KhanLFALHS2007,KermenCSJGZCMDRB2011}.  Another recent finding shows the existence of \emph{olfactory white} in human psychophysics with similar perceptual properties as white light and white audio signals \cite{WeissSYKGSS2012}.

In this paper, we investigate olfactory signals and systems at a level of abstraction removed from the physical sensing and actuation of chemical compounds.  Prior work in olfactory signal processing at the lower physical level includes the following.  One long-standing area of research has been developing chemical sensors and so-called \emph{electronic noses}, see e.g.\ \cite{GardnerB1992,PearceSNG2003,RockBW2008}, references therein and thereto.  A variety of sensing technologies including chemical gas sensors, optical sensor systems, infrared spectroscopy, and microelectromechanical sensors have been developed, and the signal processing and machine learning challenge is using raw sensor data to identify the specific composition of compounds present, cf.~\cite{Gutierrez-Osuna2002,TangCCHS2011,ZhangT2014}.  Gas chromatography mass spectrometry is considered the gold standard in laboratories, but the goal is to make portable, low-power, and low-cost systems with similar performance.  In all such systems, human perception is not considered and the goal is simply to classify according to physicochemical properties \cite{Griffin2013}.  

Moving from sensors to actuators, physical devices used for actively producing odor signals are called \emph{virtual aroma synthesizers} \cite{AdamP2013} and function by mixing compounds from several cartridges into an airstream, much like how inkjet printers produce arbitrary colors.  These devices have been put together in a variety of old and new odor communication technologies \cite{Kaye2001,Stinson2014}.  Classical examples like AromaRama and Smell-O-Vision attempted to enhance the experience of cinema viewers through a greater degree of immersion, whereas modern examples like oPhone aim to enable multi-odiferous messages transmitted to individuals. With the practice of olfactory communication, there is also an information theory of olfaction concerned with bounding the human capacity to perceive and differentiate odors \cite{HainerEJ1954,BushdidMVK02014}.

There has been much new understanding of olfactory perception and many new developments in the science of smell, see e.g.~\cite{SecundoSS2014}, which in contrast to the low-level signal processing described above, is what we build upon in this work.  An important finding is that the full gamut of odor perception for a compound or mixture of compounds (including pleasantness and whiteness) can be predicted from the physicochemical properties of the molecules \cite{SnitzYWFKS2013}, in part via information processing models of cortex \cite{StettlerA2009}.  Moreover, human olfactory perception is primarily synthetic rather than analytic.  What this means is that when people smell mixtures of compounds, they do not perceive a mixture of individual compound percepts.  Instead, they perceive a single physicochemical object all at once, where that single physicochemical object is a weighted combination of the individual physicochemical features of the compounds in the mixture.  

One may ask how the olfactory perception space is represented.  Experiments have human subjects describe the smell of pure chemical compounds in words---tolualdehyde smelling `fragrant,' `aromatic,' `almond,' and `sweet;' or valeric acid smelling `rancid,' `sweaty,' `putrid,' `fecal,' and `sickening' \cite{Dravnieks1985}---resulting in a perceptual space whose dimensions are these odor descriptors.  By averaging odor descriptor judgements over several subjects, each compound can be placed as a point in a real-valued perceptual space where the coordinate value is a function of the percentage of subjects who use a descriptor for a compound.  Such experiments have only been conducted on a small set of compounds, but we estimate the perception of uncharacterized compounds and mixtures from their physicochemical structure represented as a vector of features describing the molecule.

In particular, we learn the mapping between the physicochemical description of odorant compounds and their perceptual descriptions from a small amount of training data so it generalizes to all compounds.  We pose the learning problem as one of multivariate regression.  Our training set includes odor descriptor data (labels) and physicochemical data (features) from the small subset of compounds for which experimentally-determined odor descriptors exist.  As it is believed olfactory perception is fairly low-dimensional, we use nuclear norm regularization to keep the rank of the estimated mapping operator small \cite{LuMY2012}.

An aspect of odor perception not yet fully resolved in the literature is how perceived odor \emph{intensity} is determined by the concentration and molecular properties of compounds and the medium in which they are suspended \cite{MainlandLRL2014}.  We use straightforward concentration as an acceptable first-order approximation \cite{Cain1969}.

Given recent scientific progress on understanding olfactory perception, the time is ripe to develop engineering theories and technologies that build upon the science for applications in indoor air quality \cite{Wyon2004,VarshneyV2014a}, virtual reality \cite{Kaye2001,Stinson2014}, culinary arts\footnote{The primary contributors to human flavor perception are retronasal and orthonasal smell \cite{Shepherd2012}.} \cite{VarshneyPVSC2013,VarshneyV2014b,PinelV2014,VarshneyPVBSC2013}, hunting \cite{Sakschek1986}, and numerous others.  In this work, our contribution is to take a statistical signal processing perspective on systems involving olfaction and develop the basic tools needed to engineer them.  Upon showing how to learn the olfactory physicochemical--perceptual mapping, we develop specific example designs for problems of \emph{active odor cancellation} \cite{VarshneyV2014a} and \emph{food steganography} \cite{VarshneyV2014b}, and discuss many other olfactory operations at a high level, including filtering and smoothing, enhancement, lossy compression, communication and storage, and retrieval.  As far as we know, there is no prior work on active odor cancellation or olfactory steganography.

The remainder of this paper is organized as follows.  In Section \ref{sec:learn_mapping}, we describe the common first step for olfactory signal processing of learning the mapping between the physicochemical and perceptual spaces.  We detail the formulation of one example olfactory signal processing system, active odor cancellation, in Section \ref{sec:odor_cancel} and another, food steganography, in Section \ref{sec:food_steg}.  We provide high-level views on several other systems involving olfactory signals in Section \ref{sec:other}.  Section \ref{sec:empirical} presents empirical studies on learning, cancellation, and steganography.  Finally, Section \ref{sec:conclusion} summarizes the work and presents an outlook of future work in this area.

\section{Learning the Mapping Between Physicochemical and Perceptual Spaces}
\label{sec:learn_mapping}

The guiding principle of psychophysics, verified over centuries of experiments with human subjects, is that the physical properties of a stimulus largely determine its percept. For olfactory signals, we assume there is some general nonlinear mapping from the physicochemical attributes of a compound to its perceptual odor description.  In this section we develop a statistical methodology to learn a generalizable mapping from molecular structure of compounds to their percept.  The goal is to estimate the perceptual representation of compounds and mixtures of compounds for which no experimental ground truth on perception exists, but for which physicochemical properties are readily available. 

Human olfactory perception is difficult to pin down precisely; the most common technique used in the psychology and science literatures is to present an observer with a list of odor descriptor words or concepts and have him or her evaluate whether a given chemical's smell matches each odor descriptor.  Averaging over many individual observers yields a real-valued odor descriptor space in which each chemical compound has coordinates.  The physicochemical properties we consider are also numerical, so our goal is to learn a functional mapping between the two spaces.  In this work, we restrict ourselves to linear mappings, the validity of which is suggested by human olfaction studies \cite{KhanLFALHS2007}.  (More complex mappings, including polynomial mappings suggested in \cite{KoulakovKER2011} and kernel-based mappings \cite{SindhwaniML2013}, can be accomodated in the same type of linear model described below.)

Thus, we are given a set of training samples $\{(\mathbf{x}_1,\mathbf{y}_1), \ldots, (\mathbf{x}_n,\mathbf{y}_n)\}$ where the $\mathbf{x}_i \in \mathbb{R}^k$ are physicochemical features of compounds and the $\mathbf{y}_i \in \mathbb{R}^l$ are the perceptual vectors in the odor descriptor space.  Desiring a low-dimensional mapping, we use nuclear norm-regularized multivariate linear regression to learn a matrix $\mathbf{A}^* \in \mathbb{R}^{l\times k}$ that maps unseen compounds from the chemical to the perceptual space.  In particular, if we concatenate all the training samples into matrices $\mathbf{X} \in \mathbb{R}^{k\times n}$ and $\mathbf{Y} \in \mathbb{R}^{l\times n}$, the problem to solve is:
\begin{equation}
\label{eq:regression}
	\mathbf{A}^* = \arg\min_{\mathbf{A}} \|\mathbf{Y} - \mathbf{A}\mathbf{X}\|_F + \lambda \|\mathbf{A}\|_{*}
\end{equation}
where $\lambda$ trades data fidelity for sparsity of the singular values of $\mathbf{A}^*$.  This problem is convex and can be solved by interior point methods and a variant of Nesterov's smooth method \cite{LuMY2012}.  

Note that Euclidean norms make sense as both optimization objectives (Frobenius norm) and characterizations of system performance (RMSE), since they are used in olfactory psychophysics studies with human subjects from several different laboratories \cite{KoulakovKER2011, CastroRC2013, SnitzYWFKS2013, YanLF2015}, and in the recent DREAM Olfaction Prediction Challenge, a part of the Rockefeller University Smell Study.\footnote{{\tt https://www.synapse.org/\#!Synapse:syn2811262}}

\section{Active Odor Cancellation}
\label{sec:odor_cancel}

Noise cancellation is one of the most basic of signal processing tasks \cite{Wiener1949,BlackmanBS1946}, and thus we use it as the first task within which to describe olfactory signal processing.  There are often settings where chemical signals should be canceled: poor indoor air quality and malodors are not only a nuisance and source of dissatisfaction, but can decrease the productivity of office workers six to nine percent \cite{Wyon2004}.  Four general categories of techniques are currently used for reducing or eliminating odors: \emph{masking}, which attempts to `overpower' the offending odor with a single pleasant odor; \emph{absorbing}, which uses active ingredients like baking soda and activated carbon; \emph{eliminating}, in which chemicals react with odor molecules to turn them into inert, odorless compounds; and \emph{oxidizing}, which accelerates the break-down of malodorous compounds.  Instead, here we develop a statistical signal processing method for performing active odor cancellation, with some resemblance to active noise and vibration cancellation \cite{BoseC1984,Brown2005}.

We approach the problem by taking advantage of the psychophysical properties of human end-consumers of odor.  In particular, there is a recently discovered percept called olfactory white, which is the neutral smell generated by equal-intensity stimuli well-distributed across the physicochemical space \cite{WeissSYKGSS2012}, much like white light or auditory noise.  More specifically, the set of all odorant compounds spans a particular subspace of physicochemical attribute values; when more than approximately thirty compounds, all diluted to a concentration having equal perceived intensity, are smelled together in a mixture, the resulting percept is a neutral odor that is the same no matter which compounds are included in the mixture, but only if the compounds in the set are fairly uniformly distributed in the subspace \cite{WeissSYKGSS2012}.  Whiteness is a central concept in active signal cancellation generally speaking \cite{BlackmanBS1946}.  Our goal in this section is to sense an existing malodor and to output a compound mixture from a virtual aroma synthesizer such that the resulting combined odor is white.

In the active odor cancellation applications of interest to us, several different malodors may be sensed and canceled by the same virtual aroma synthesizer.  Therefore, in addition to providing excellent cancellation performance, we also desire the cardinality of the compound set in the system to be minimized because it is costly to have many cartridges.  Toward this goal, we use the group lasso or simultaneous sparsity-inducing $\ell_1/\ell_2$ norm \cite{YuanL2006}.  We also require a non-negativity constraint because optimized compound mixtures can only be output into the air, not subtracted \cite{KimMP2012}.  Due to the synthetic nature of human olfaction, the generally nonlinear perceptual mapping (simplified to linear in this paper) is applied to the physicochemical representation of mixtures of compounds exhaled by the system.

As a starting point, we collect a set of $n$ compounds that could possibly be used in the aroma synthesizer.  Let the physicochemical representation of this dictionary be $\mathbf{X}_{\text{dict}} \in \mathbb{R}^{k\times n}$.  We would like to design the system to optimally cancel $m$ different malodors with perceptual representations $\mathbf{Y}_{\text{mal}} \in \mathbb{R}^{l \times m}$.  We would like to determine a simultaneously sparse set of non-negative coefficients $\mathbf{W}^* \in \mathbb{R}^{n\times m}_{+}$ that minimize: 
\begin{equation}
\label{eq:odor_cancel}
	\tfrac{1}{2}\|\mathbf{Y}_{\text{mal}} + \mathbf{A}^*\left(\mathbf{X}_{\text{dict}}\mathbf{W}\right)\|_F^2 + \mu\|\mathbf{W}\|_{1,2}, \quad \text{s. t. } \mathbf{W} \ge 0,
\end{equation}
where $\mu$ is a regularization parameter, and the $\ell_1/\ell_2$ norm takes $\ell_2$ norms of each of the $n$ length-$m$ rows of $\mathbf{W}$ first and then takes the $\ell_1$ norm of the resulting length-$n$ vector.  The physicochemical-to-perceptual mapping $\mathbf{A}^*$ comes from the learning problem described in Section~\ref{sec:learn_mapping}.  The data fidelity term is $\|\mathbf{Y}_{\text{mal}} + \mathbf{A}^*\left(\mathbf{X}_{\text{dict}}\mathbf{W}\right)\|_F^2$ because the all-zeros vector in the perceptual space is an olfactory white.

A more general formulation for the optimization can incorporate the fact that there is not a single olfactory white, but a family of them.  The more general optimization problem is then a minimization over the coefficients and the particular olfactory whites in the family:
\begin{equation}
\label{eq:odor_cancel_general}
	\tfrac{1}{2}\|\mathbf{Y}_{\text{mal}} + \mathbf{A}^*\left(\mathbf{X}_{\text{dict}}\mathbf{W}\right) - \mathbf{Y}_{\text{white}}\|_F^2 + \mu\|\mathbf{W}\|_{1,2}, \quad \text{s. t. } \mathbf{W} \ge 0, \, \mathbf{Y}_{\text{white}} \in \mathcal{Y}_{\text{white}}.
\end{equation}
The family $\mathcal{Y}_{\text{white}}$ can be specified as the set of matrices with all rows within a column having the same value, i.e., of the form $\boldsymbol{1}\mathbf{c}^T$, where $\boldsymbol{1}$ is the length $l$ all-ones vector and $\mathbf{c}$ is a length $m$ vector that is free to be decided upon.

\section{Food Steganography}
\label{sec:food_steg}

Many children (and adults) are picky eaters to whom junk food is more attractive than healthy food.  This instinct was useful for hunter-gatherers that depended heavily on their senses to decide what to eat: in nature, sweet foods are almost always safe and nutritious whereas foods that smell odd are potentially toxic or spoiled.  In modern environments, this same instinct often serves to make people obese and chronically ill. Hiding a nutritious, averse food in a delectable food may therefore aid people in eating healthier.  

\begin{figure}
  \centering
  \includegraphics[width=\columnwidth]{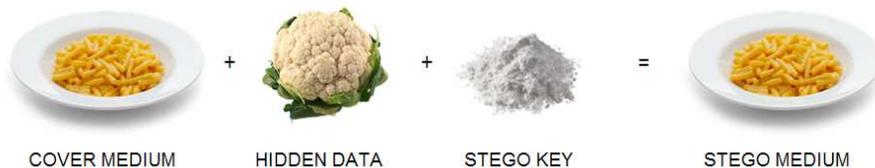}
  \caption{Depiction of food steganography in the perceptual domain, where macaroni \& cheese is delectable, cauliflower is averse, and the white powder is the additive.}
  \label{fig:stego_eq}
\end{figure}

A second problem, closely related to active odor cancellation, is hiding the flavor of one food inside the flavor of another food through the use of an additive: \emph{food steganography}.  Steganography is the very old concept of imperceptibly hiding a signal into a cover medium \cite{BenderGML1996,JohnsonJ1998,ProvosH2003,ZielinskaMS2014}, which has a signal processing flavor in approaches like spread spectrum image steganography \cite{MarvelBR1999}.  We demonstrate a statistical signal processing approach to optimally design a food additive (either using pure compounds or natural ingredients) to act as a steganographic key for this food steganography problem.  The steganographic percepts are depicted in Fig.~\ref{fig:stego_eq}, illustrated using the hiding of cauliflower inside of macaroni and cheese as an example. Note that there are many possible goals in steganography; herein the goal is not for the receiver to decipher a hidden message, but only to make imperceptible a food to which the receiver is averse (and which may have good nutritional properties).

A food additive (steganographic key) combines with the averse food (hidden signal), and the delectable food (cover medium) such that the combination is perceived as only the delectable food's flavor; the olfactory white signal is used as a mathematical intermediary.  The food additive may be composed of some weighted mixture of pure compounds or some weighted mixture of food ingredients from a dictionary.  We may also want to regularize the problem by including a sparsity or other cost-related penalty on the food additive.
 
There are typically tens to hundreds of different chemical compounds contributing to flavor per food ingredient.  Using data on the concentrations of compounds in foods, we take a weighted combination of the physicochemical vectors of the constituent compounds of a food to determine its perceptual representation using the mapping learned in Section~\ref{sec:learn_mapping}.  Next, we solve a regularized inverse problem with a non-negativity constraint to find compounds or foods and their coefficients required to synthesize an additive that produces olfactory white when combined with an averse food of interest.  

Let $\mathbf{X}_{\text{cov}}$ be the physicochemical representation of the cover medium's compounds and $\mathbf{w}_{\text{cov}}$ be the concentrations of the cover medium's compounds.  Likewise let us introduce $\mathbf{X}_{\text{hid}}$ and $\mathbf{w}_{\text{hid}}$ for the hidden data.  Let $\mathbf{X}_{\text{dict}}$ be a dictionary of $n$ possible compounds from which we can construct the steganographic key (food additive) along with its weight vector $\mathbf{w}_{\text{dict}}$, which is the subject of design.    First, with a general nonlinear physicochemical-to-perceptual mapping $A^*(\cdot)$, the perceptual hiding we want to perform is to choose $\mathbf{w}_{\text{dict}}$ to satisfy:
\begin{equation}
	A^*\left(\mathbf{X}_{\text{cov}}\mathbf{w}_{\text{cov}} + \mathbf{X}_{\text{hid}}\mathbf{w}_{\text{hid}}  + \mathbf{X}_{\text{dict}}\mathbf{w}_{\text{dict}}\right) \approx A^*\left(\mathbf{X}_{\text{cov}}\mathbf{w}_{\text{cov}}\right).
\end{equation}

With the linear mapping that we are assuming in this work, the objective simplifies to:
\begin{equation}
	\mathbf{A}^*\mathbf{X}_{\text{hid}}\mathbf{w}_{\text{hid}} \approx -\mathbf{A}^*\mathbf{X}_{\text{dict}}\mathbf{w}_{\text{dict}}.
\end{equation}
The determination of the steganographic key does not depend on the cover medium, and is simply the odor cancellation signal from Section~\ref{sec:odor_cancel}.  Specifically, we solve the following optimization problem:
\begin{align}
\label{eq:inverse}
&\min_{\mathbf{w}_{\text{dict}}} \|\mathbf{A}^*\mathbf{X}_{\text{hid}}\mathbf{w}_{\text{hid}} + \mathbf{A}^*\mathbf{X}_{\text{dict}}\mathbf{w}_{\text{dict}}\|_2^2 + \nu J\left(\mathbf{w}_{\text{dict}}\right) \\ \notag
&\ \text{s. t. }\mathbf{w}_{\text{dict}} \ge \mathbf{0}
\end{align}
where $J(\cdot)$ could be one of a number of possible regularization terms meant to promote secondary objectives such as monetary frugality, sparsity, or nutrition.

In the case that we only want to use a set of $n'$ food ingredients to compose the additive, we first use data on the known concentrations of compounds in food ingredients to construct an $n \times n'$ weight matrix $\mathbf{W}_{\text{ingr}}$ that multiplies $\mathbf{X}_{\text{dict}}$ to obtain a dictionary of food ingredient physicochemical features.  The weight vector to be solved for is then an $n' \times 1$ vector $\mathbf{w}_{\text{ingr}}$:
\begin{align}
\label{eq:inversefood}
&\min_{\mathbf{w}_{\text{ingr}}} \|\mathbf{A}^*\mathbf{X}_{\text{hid}}\mathbf{w}_{\text{hid}} + \mathbf{A}^*\mathbf{X}_{\text{dict}}\mathbf{W}_{\text{ingr}}\mathbf{w}_{\text{ingr}}\|_2^2 + \nu J\left(\mathbf{w}_{\text{ingr}}\right) \\ \notag
&\ \text{s. t. }\mathbf{w}_{\text{ingr}} \ge \mathbf{0}.
\end{align}

\section{Other Systems}
\label{sec:other}

We have detailed two specific olfactory signal processing systems in the previous two sections.  However, the variety of possible tasks one may want to perform in the olfactory modality is as broad as in other modalities.  In this section, we discuss formulations for several of those other possible tasks.  This section is not meant to be an exhaustive coverage of all possible olfactory signal processing tasks, but is meant to showcase the realm of possibilities and spur future research.

The main terms in the objectives of cancellation and steganography \eqref{eq:odor_cancel}, \eqref{eq:inverse}, and \eqref{eq:inversefood} are as they are because we desire a neutral all-zeros perception as output in those tasks.  However, there is nothing preventing us from inserting a desired target or output odor percept into the objective (as we did for the full family of olfactory whites), which would allow us to perform general filtering or equalization operations.  With a given desired perceptual output $\mathbf{y}_{\text{des}}$, the problem is:
\begin{equation}
\label{eq:general_static}
	\min_{\mathbf{w}} \|\mathbf{A}^*\mathbf{x}_{\text{in}} + \mathbf{A}^*\mathbf{X}_{\text{dict}}\mathbf{w} - \mathbf{y}_{\text{des}}\|_2^2 + \mu J\left(\mathbf{w}\right), \text{ s. t. } \mathbf{w} \ge 0,
\end{equation}

Olfactory cancellation, filtering, or equalization will, in general, take place in dynamic rather than static environments.  For example, think of the indoor air quality of an automobile traveling from a chemical plant to an urban environment via a garbage dump.  Problem \eqref{eq:general_static} can be extended to include a time variable $t$ that applies to $\mathbf{x}_{\text{in}}$, $\mathbf{w}$, and possibly $\mathbf{y}_{\text{des}}$.  The dynamic version of the problem can be addressed using the theory of adaptive filtering; one specific way to formulate the adaptive version is through a variation on a regularized LMS algorithm with nonnegativity constraint \cite{ChenGH2010,ChenRBH2011,ChenRBH2014}.  

The difference from the standard adaptive linear combiner here is that $\mathbf{w}_t$ is multiplied by the dictionary of available compounds and $\mathbf{x}_{\text{in},t}$ is an additive term; $\mathbf{w}_t$ and $\mathbf{x}_{\text{in},t}$ are not multiplied or convolved with each other.  The update rule for an LMS-like adaptive filter is:
\begin{multline}
\label{eq:LMS_update}
	\mathbf{w}_{t+1} = \mathbf{w}_t - 2\eta\diag\{\mathbf{w}_t\} \\\times \left((\mathbf{A}^*\mathbf{X}_{\text{dict}})^T(\mathbf{A}^*\mathbf{x}_{\text{in},t} + \mathbf{A}^*\mathbf{X}_{\text{dict}}\mathbf{w}_t - \mathbf{y}_{\text{des},t})  + \mu\partial J(\mathbf{w}_t)\right),
\end{multline}
where $\eta$ is the step size of the LMS algorithm.

Virtual reality applications may require a `smelltrack' similar to a soundtrack to accompany motion pictures.  In such applications, we can assume that the desired olfactory perception signal over the entire time period is known in advance and that there is no ambient odor to overcome.  Therefore, we may be able to do better than adaptive filtering.  The difficulty is that compounds exhaled into an environment linger for some duration.  If we develop a stochastic model for this persistence, perhaps using a Gaussian puff model \cite{TerejanuSS2007,FoxFW2007} composed with a psychophysical sniff model~\cite{VerhagenWNWW2007,MainlandS2006}, then we can use appropriate extensions of the Rauch-Tung-Striebel smoother to obtain an optimal control strategy of virtual aroma synthesizer actuation.  Such an approach can also allow the composer to only specify `key frames' of smell with the signal processing algorithm interpolating the rest.

An alternative to specifying a desired output signal $\mathbf{y}_{\text{des}}$, as in adaptive filtering, is to specify the desired behavior and requirements of the filter to be designed without using a specific input signal realization in the objective \cite{WeiSO2013}.  For example, a desired behavior might be to allow pleasant odor components to pass through the filter unchanged and to cancel unpleasant odor components (similar to a low-pass filter), or to allow all odors except for the odor descriptor `vomit' to pass (similar to a notch filter).  Such filters do not depend on the input signal, but have the same multiplicative or convolutive behavior for all inputs.  

Unfortunately in our signal processing approach, we affect the synthesized perceptual representation of the output through the \emph{superposition} of the input and a set of compounds we design, not by physically filtering different compounds or types of compounds.  This means, as previously noted, that the specification $\mathbf{w}$ does not multiply (or convolve) the input signal $\mathbf{x}_{\text{in}}$, but adds to it after being modulated by the dictionary $\mathbf{X}_{\text{dict}}$.  This implies filters implementing multiplicative behavior are not possible for olfactory signals.  Letting $\mathbf{y}_{\text{in}} = \mathbf{A}^*\mathbf{x}_{\text{in}}$ and $\mathbf{y}_{\mathbf{w}} = \mathbf{A}^*\mathbf{X}_{\text{dict}}\mathbf{w}$, we are saying that it is not possible to choose a $\mathbf{y}_{\mathbf{w}}$ such that $\mathbf{y}_{\text{in}} + \mathbf{y}_{\mathbf{w}} = \mathbf{0}$ for all $\mathbf{y}_{\text{in}}$.  

Furthermore, we cannot define general filterbank decompositions either.  What is not precluded, however, is decomposing specific realizations of odor signals.  For example, extending the idea of Kisstixx lip balm,\footnote{{\tt http://kisstixx.com/ }} we can decompose a well-recognized perceived odor into two different perceived odors which are each well-recognized separately, just like decomposing into parts with non-negative matrix factorization \cite{LeeS1999}.

The learned perceptual mapping $\mathbf{A}^*$ allows us to define a distortion function between compound mixtures that can be used for a variety of processing tasks including lossy compression (i.e. choosing a different, less costly set of compounds to approximately reproduce the odor of the original set of compounds), denoising, hashing, and retrieval.  Moreover, the storage and communication of olfactory signal data only needs to be in the synthetic perceptual domain; data on the analytic physical compounds of smells is not required.  

Going one step further, since pleasantness is the main component of odor signals, storing and communicating only the pleasantness scalar value or scalar time series is often the only desire \cite{CheeSV2013}.  For example, pleasantness and not any higher-order odor dimensions, is one of the two main criteria in the selection step of a successful computational creativity system for culinary recipes \cite{VarshneyPVSC2013,VarshneyPVBSC2013,PinelV2014}.

Finally, we conjecture that odor enhancement can be approached in much the same way as image enhancement via hue and saturation.  As we noted earlier, olfactory white emerges only with a large number of compounds that span the perceptual space.  Olfactory white can be viewed as a fully desaturated odor.  This reasoning implies that we can start with an odor and decrease its saturation by gradually adding many compounds from different parts of the space, and increase its saturation by adding to the concentration of a few compounds from one part of the space.  We can change the hue of an odor, e.g.\ start with the smell of a rose and make it `fishier' \cite{PinelSV2014}, by adding appropriate compounds while ensuring not to alter the saturation.

\section{Empirical Results}
\label{sec:empirical}

In this section, we present illustrative empirical results on learning the mapping from physicochemical features to odor descriptors and using this mapping for active odor cancellation in a setting that may arise in the break room or lunch room of a small office.  We also illustrate the use of this mapping for food steganography where the hidden food is cooked broccoli, which has many positive nutritional qualities but to which many have aversion.

\subsection{Learning the Mapping}
\label{sec:empirical:learn_mapping}

The first step in our empirical study is to learn the mapping $\mathbf{A}^*$ from physicochemical properties of compounds to the olfactory perception of those compounds.  We collect a ($k = 18$)-dimensional physicochemical feature vector for each of 143 different chemical compounds that have been judged by human observers against $l = 146$ different odor descriptors as diverse as `almond,' `cat urine,' `soapy,' `stale tobacco smoke,' and `violets.'  The 18 physicochemical features are obtained from the National Center for Biotechnology Information's PubChem Project and include among others: topological polar surface area, partition coefficient prediction (XLogP), molecular weight, complexity, heavy atom count, hydrogen bond donor count, and tautomer count.  The feature values are the properties exhibited by a single molecule of the compound, e.g., for ethyl pyrazine, topological polar surface area = 25.8, XLogP = 0.7, molecular weight = 108.14112, and so on.  The human judgements on odor descriptors are obtained from the Atlas of Odor Character Profiles (AOCP) \cite{Dravnieks1985}, which pools data from a panel of hundreds of flavor/fragrance experts.\footnote{We use the percentage of applicability data from AOCP.  For each compound, many human experts evaluate it against each of the 146 descriptors on a zero to five scale.  The percentage of applicability for a given descriptor ranges from zero to one hundred and is the geometric mean of the percentage of experts who give a score greater than zero and the ratio of the sum of the scores and five times the number of experts as a percentage \cite{Dravnieks1985}.  When interpreting accuracy results in this section, the definition of percentage of applicability should be kept in mind.} The two data sets are matched and joined using Chemical Abstracts Service (CAS) Registry numbers.

\begin{figure}
  \centering
  \includegraphics[width=0.65\textwidth]{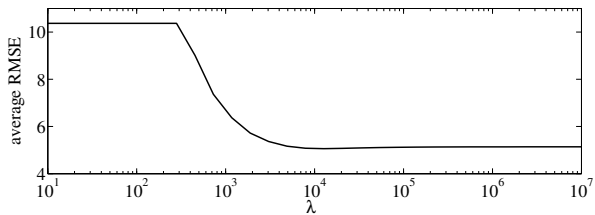}
  \caption{Five-fold cross-validation testing root mean squared error of the mapping between physicochemical and perceptual 
	spaces averaged across the 146 perceptual dimensions.}
  \label{fig:learn_mapping_cv}
\end{figure}

Using the percept matrix $\mathbf{Y} \in \mathbb{R}^{146 \times 143}$ from AOCP and the physicochemical matrix $\mathbf{X} \in \mathbb{R}^{18\times 143}$ from PubChem, we learn the mapping by solving the nuclear norm-regularized multivariate linear regression problem discussed in Section~\ref{sec:learn_mapping} using the method of \cite{LuMY2012}.  We conduct five-fold cross-validation to determine the best value of $\lambda$.  As a figure of merit, we consider the root mean squared error (RMSE) averaged over the 146 dimensions; Fig.~\ref{fig:learn_mapping_cv} shows the cross-validation testing average RMSE as a function of $\lambda$.  
The error is minimized at approximately $\lambda = 10^4$ and is the value we use going forward.

\subsection{Active Odor Cancellation}
\label{sec:empirical:odor_cancel}

We consider $m = 4$ different offending odors that we wish to cancel with the same, small-cardinality set of olfactory compounds.  The four smells are: durian (Durio zibethinus), onion (Allium cepa L.), katsuobushi (dried bonito), and sauerkraut. With an optimal solution to the problem \eqref{eq:odor_cancel}, we can create a device with minimal complexity that senses the current odor and outputs the appropriate concentrations of compounds to cancel it.  When placed in a lunch room, the device will be able to cancel these four odors, but also many others.

\begin{figure}
  \centering
  \includegraphics[width=\textwidth]{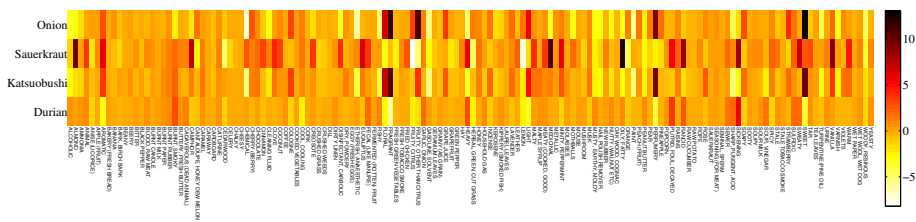}
  \caption{Perceptual projection of the mixture of compounds contained in durian, katsuobushi, sauerkraut, and onion.}
  \label{fig:fourpercept}
\end{figure}

The perceptual representation of the four odor mixtures of interest can be predicted from the learned mapping.  First, in the same spirit as the synthesis that takes place in human olfactory perception, we take a linear combination of the physicochemical features of the components of the odor and then map the resulting physicochemical vector to perceptual space.  We obtain the set of olfactory compounds present in the four odors and their concentrations from the Volatile Compounds in Food 14.1 database (VCF) and obtain physicochemical features of those compounds from PubChem.  The resulting predicted perceptions of durian, katsuobushi, sauerkraut, and onion are shown in Fig.~\ref{fig:fourpercept}.  
For example, it can be seen in the figure that sauerkraut is perceived most like the `oily, fatty' descriptor and least like the `fruity, citrus' descriptor.  The odor descriptors with largest positive coefficients for the other three malodors are `sickening,' `fragrant,' and `sweet,' respectively.  The odor descriptors with largest negative coefficients for the other three malodors are `fragrant,' `chemical,' and `chemical,' respectively.

Having predicted the perception of the four odors of interest, the next step is to find compounds that can be used to cancel their smells perceptually.  Toward this end, we first construct a dictionary of compounds from which we can find the cancellation set.  We extract $n = 5736$ compounds from VCF found naturally in food and find their physicochemical properties from PubChem.  This dictionary, with members only from natural edible products has certain limitations, which we comment on later.  We use the non-negative simultaneous sparsity formulation given in Section~\ref{sec:odor_cancel} with this dictionary to find the optimal sparse set of compounds for active odor cancellation with different values of the regularization parameter $\mu$.  We use SDPT3 to solve the optimization problem \cite{TohTT1999}.

\begin{figure}
  \centering
  \includegraphics[width=0.65\textwidth]{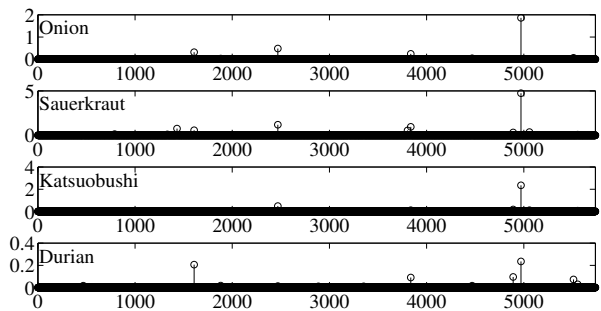}
  \caption{Dictionary coefficient values in optimal cancellation solution with $\mu = 1$.}
  \label{fig:coeff_stem_1}
\end{figure}
\begin{figure}
  \centering
  \includegraphics[width=\textwidth]{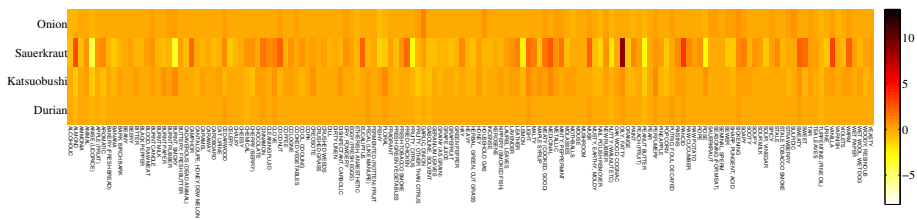}
  \caption{Perceptual representation of residual odor after cancellation of durian, katsuobushi, sauerkraut, and onion with $\mu = 1$.}
  \label{fig:fourpercept_residual_1}
\end{figure}
\begin{figure}
  \centering
  \includegraphics[width=\textwidth]{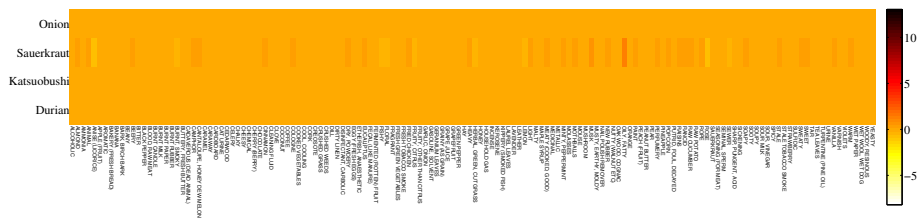}
  \caption{Perceptual representation of residual odor after cancellation of durian, katsuobushi, sauerkraut, and onion with $\mu = 0.25$.}
  \label{fig:fourpercept_residual_p25}
\end{figure}

The set of coefficients $\mathbf{W}$ found for $\mu = 1$ is shown in Fig.~\ref{fig:coeff_stem_1}.  There are 22 compounds with positive coefficient value in at least one of the four cancellation additives.  The residual odor remaining after cancellation is shown in Fig.~\ref{fig:fourpercept_residual_1}.  
\begin{figure}
  \centering
  \includegraphics[width=0.65\textwidth]{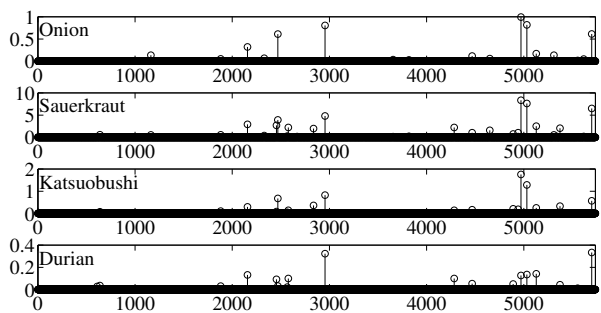}
  \caption{Dictionary coefficient values in optimal cancellation solution with $\mu = 0.25$.}
  \label{fig:coeff_stem_p25}
\end{figure}
The Frobenius norm of the residual is 17.13 and the $\ell_2$ norms of the individual odors are 1.41 for durian, 4.38 for katsuobushi, 16.30 for sauerkraut, and 2.50 for onion.  By reducing $\mu$, we can improve the cancellation at the expense of increasing the number of compounds used.  The coefficients in the optimal solution for $\mu = 0.25$ are shown in Fig.~\ref{fig:coeff_stem_p25} and the residual perception in Fig.~\ref{fig:fourpercept_residual_p25}.
In this solution, 38 compounds have positive coefficients and the Frobenius norm of the residual is 2.30.  Residual $\ell_2$ norms of individual odors are: durian 0.04, katsuobushi 0.12, sauerkraut 2.29, and onion 0.24.

The $\mu = 1$ solution does provide a certain level of odor cancellation, but just by decreasing the sparsity a little bit, we are able to get very good cancellation.  Only the residual of sauerkraut is non-negligible in the $\mu = 0.25$ solution, and even that is nearing negligibility.  We note that certain parts of the various odor signatures are easier to cancel than others.  For example, the descriptor `medicinal' is mostly removed from the sauerkraut solution with $\mu = 1$ but `eucalyptus' is not.  With a limited budget on their number, compounds that affect all four odors are at a premium.  Thirteen compounds (out of 22 and 38, respectively) are common to the two solutions: `(+)-cyclosativene,' `(E,E,Z)-1,3,5,8-undecatetraene,' `(R)-3-hydroxy-2-pentanone,' `1,3,5,8-undecatetraene,' `10-methyl-2-undecenal,' `cis-piperitol oxide,' `cubenene,' `cyclooctatetraene,' `dehydrocurdione,' `ethylpyrrole (unkn.str.),' `heptatriacontane,' `juniper camphor,' and `methane.'

\begin{figure}
  \centering
  \includegraphics[width=0.65\textwidth]{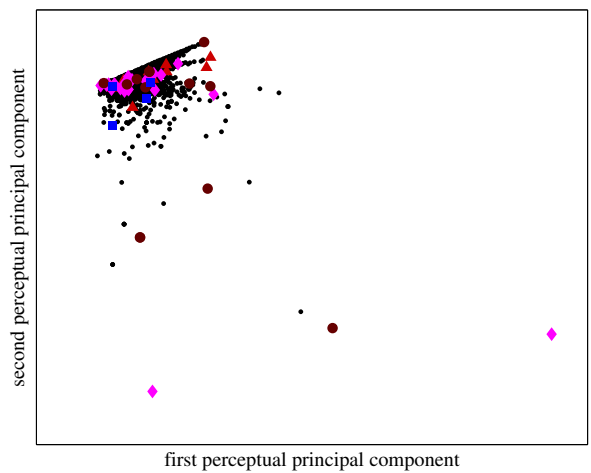}
  \caption{Principal component projection of perceptual vectors of dictionary and four odors.  The blue squares are the four odors to be canceled, the red triangles are compounds selected only in the $\mu = 1$ solution, the magenta diamonds are compounds selected only in the $\mu = 0.25$ solution, the maroon circles are the compounds selected in both the $\mu = 1$ and $\mu = 0.25$ solutions, and the black points are all other compounds in the dictionary.}
  \label{fig:four_odor_perceptual_pca}
\end{figure}

As discussed in Section~\ref{sec:introduction}, our formulation of active odor cancellation is associated with the concept of olfactory white, which emerges with around thirty (but not with fewer) compounds of equal intensity covering the space of compounds fairly evenly.  We visualize the space of compounds using the first two principal components of the perceptual vectors of the compounds in the dictionary and the four odors under consideration in Fig.~\ref{fig:four_odor_perceptual_pca}.
The compounds with non-zero coefficient values do span the space as best as they can to produce something akin to olfactory white.  It is interesting to note that the modest increase from 22 to 38 compounds yields such a large improvement in cancellation quality where these two values are on either side of the number required for olfactory white.  In the visualization, we also see that the dictionary we have used does not well-cover the full space; this is partly because the only compounds we have used are present in food products, suggesting that for improved cancellation, we should consider a more diverse dictionary that covers the space of olfactory perception better.

\subsection{Food Steganography}
\label{sec:empirical:food_steg}

\begin{figure*}
  \centering
  \includegraphics[width=\textwidth]{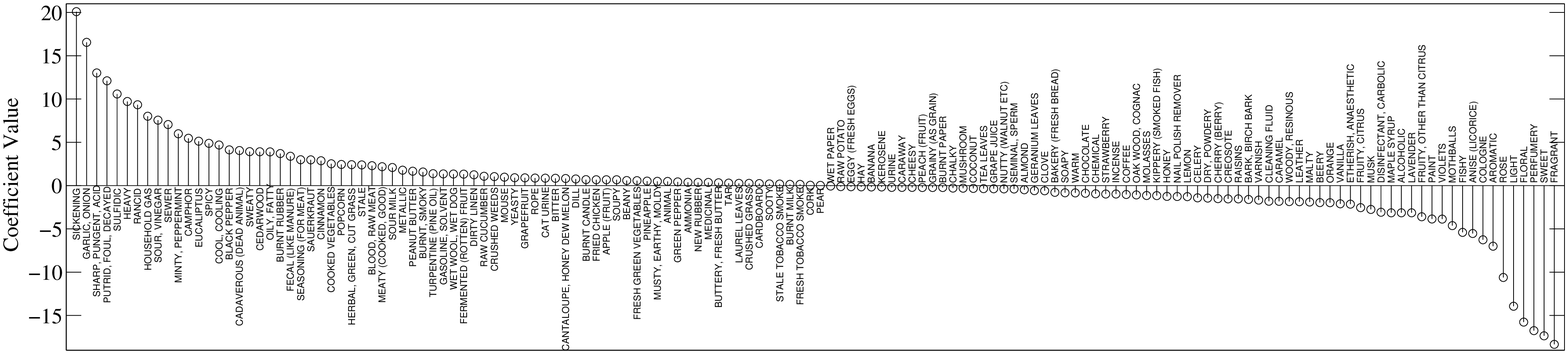}
  \caption{Perceptual projection of the mixture of compounds contained in broccoli.}
  \label{fig:broccoli_percept}
\end{figure*}

To demonstrate our approach to food steganography, we design food additives to act as steganographic keys for cooked broccoli, where the cover medium may be cheese or mango juice.  (As discussed in Section~\ref{sec:food_steg}, the cover medium does not matter under the linearity assumption.)  Similarly to durian, katsuobushi, sauerkraut, and onion, we first characterize broccoli physicochemically and perceptually.  The $21$ compounds in cooked broccoli from VCF are given in Table~\ref{table:broccoli_compounds}.
\begin{table}
	\begin{center}
	\begin{tabular}{|c|l|} \hline
Conc. & Compound Name \\\hline
0.0065 & benzaldehyde \\\hline
0.0324 & 1-octanol \\\hline
0.0162 & 4-methylacetophenone \\\hline
0.0811 & phenylacetaldehyde (=benzeneacetaldehyde) \\\hline
0.2596 & nonanal (=pelargonaldehyde) \\\hline
0.0162 & limonene \\\hline
0.0973 & phenethyl isothiocyanate \\\hline
0.0162 & (E,E)-2,4-decadienal \\\hline
0.0649 & dimethyl trisulfide (=2,3,4-trithiapentane, methyltrithiomethane) \\\hline
0.0162 & 2-pentylfuran \\\hline
0.0162 & 2,3,5-trithiahexane \\\hline
0.0162 & (E,Z)-2,4-heptadienal \\\hline
0.0973 & (E,E)-2,4-heptadienal \\\hline
0.4867 & 4-(methylthio)butyl isothiocyanate \\\hline
0.0162 & 2-hexenal \\\hline
0.6489 & 5-(methylthio)pentanenitrile \\\hline
0.0162 & dimethyl disulfide (=methyldithiomethane) \\\hline
0.4867 & 3-phenylpropanenitrile (=phenethyl cyanide, benzenepropanenitrile) \\\hline
0.0227 & 1,2-dimethoxybenzene (=veratrole) \\\hline
0.0649 & (Z)-3-hexen-1-ol (=leaf alcohol) \\\hline
0.0162 & benzothiazole \\\hline
	\end{tabular}
	\end{center}
	\caption{Compounds in cooked broccoli with concentrations.}
	\label{table:broccoli_compounds}
\end{table}
We take the concentration values as the weights $w_j$ and normalize to unit $\ell_2$ norm, obtaining the physicochemical representation of the mixture $\mathbf{X}_{\text{hid}}\mathbf{w}_{\text{hid}}$.  The result of projecting the mixture into perceptual space using $\mathbf{A}^*$ is shown in Fig.~\ref{fig:broccoli_percept}.  
The most prominent predicted odor descriptors for cooked broccoli are sickening, garlic/onion, and sharp/pungent/acid, which speaks to why many people dislike it.

The pure compounds dictionary associated with inverse problem \eqref{eq:inverse} is the same one used in the active odor cancellation study of compounds found naturally in food with $n = 5736$.  We also construct an $n' = 297$ food ingredients dictionary from VCF data associated with the inverse problem \eqref{eq:inversefood}.  Specifically, we only include food products with at least 15\% of their listed compounds having both a match in PubChem and having a concentration value listed.  If a range of concentrations is listed in VCF we use the midpoint of the range; if the value is listed as `trace,' we use the value $10^{-6}$ parts per million.  All food ingredient concentrations are normalized to have unit $\ell_2$ norms.  The union of compounds found in the 297 food ingredients is a subset of the 5736 compounds in the pure compound dictionary.

We use a sparsity-promoting penalty for $J$ as a demonstration.  The result based on the pure compound dictionary is shown in Table~\ref{table:broccoli_additivecomp} and the result based on the food ingredient dictionary is shown in Table~\ref{table:broccoli_additivefood}. 
\begin{table}
	\begin{center}
	\begin{tabular}{|r|l|} \hline
Conc.~ & Compound \\\hline
10.4520 & methane \\\hline
5.6617 & 2,5-hexanedione (=acetonylacetone) \\\hline
4.6890 & cyclotetracosane \\\hline
3.1862 & cubenene \\\hline
1.7275 & 1,1'-dioxybis(1-decanol) \\\hline
0.6456 & 2,4-diphenylpyrrole \\\hline
0.5931 & propanamide \\\hline
0.5685 & cyclooctatetraene \\\hline
0.5044 & heptatriacontene (unkn.str.) \\\hline
0.3386 & p-1,5-menthadien-7-ol \\\hline
0.3376 & 2-ethyl-5-pentanoylthiophene \\\hline
0.1209 & ethylpyrrole (unkn.str.) \\\hline
0.1106 & docosahexaenoic acid (unkn.str.) \\\hline
0.0224 & 10-methyl-2-undecenal \\\hline
0.0055 & $\alpha$-maaliene \\\hline
0.0041 & 2-(2-methylbutanoyl)furan \\\hline
	\end{tabular}
	\end{center}
	\caption{Additive mixture composed of pure compounds for food steganography with cooked broccoli as the hidden data.}
	\label{table:broccoli_additivecomp}
\end{table}
\begin{table}
	\begin{center}
	\begin{tabular}{|r|l|} \hline
Conc.~ & Food Product Name \\\hline
13.2999 & ANGELICA SEED OIL \\\hline
7.5619 & CUMIN SEED (Cuminum cyminum L.) \\\hline
7.5328 & MUSSEL \\\hline
4.3985 & BARLEY (unprocessed) \\\hline
2.8275 & LOBSTER \\\hline
2.7808 & BLACKBERRY BRANDY \\\hline
2.5717 & ROSE WINE \\\hline
2.3048 & OTHER VITIS SPECIES \\\hline
1.4727 & TURNIP \\\hline
1.3033 & LAMB and MUTTON FAT (heated) \\\hline
0.8432 & INDIAN DILL ROOT (Anethum sowa Roxb.) \\\hline
0.6520 & LOGANBERRY (Rubus ursinus var. loganobaccus) \\\hline
0.4794 & ELDERBERRY FRUIT \\\hline
0.1626 & PEANUT (raw) \\\hline
0.0989 & MICROCITRUS SPECIES OIL \\\hline
0.0285 & PRAWN \\\hline
	\end{tabular}
	\end{center}
	\caption{Additive mixture composed of food ingredients for food steganography with cooked broccoli as the hidden data.}
	\label{table:broccoli_additivefood}
\end{table}
It is difficult to interpret the pure compounds solution.  The food ingredient solution is easier to interpret.  Angelica seeds, which are the main component of the food product-based additive, have a very unique pleasant smell entirely unlike similar plants such as fennel, parsley, anise, and caraway, and are used as a flavoring in Scandinavian cuisine.

\section{Conclusion}
\label{sec:conclusion}

This paper represents a first foray for statistical signal processing into the new multimedia domain of human olfaction, building on new developments in the science of smell.  The general framework was demonstrated through two specific applications in active odor cancellation and in food steganography, and methods for solving a broader class of problems were also indicated.  Empirical results from the design procedures required bringing together data on the flavor composition of ingredients (from gas chromatography--mass spectrometry), the molecular properties of odor compounds (from chemoinformatics), and the human perception of flavors (from hedonic psychophysics) with algorithmic techniques for function learning and inverse problem solution.

By addressing one of the fundamental problems of signal processing, noise cancellation, this work opens up a new category of techniques for dealing with bad odors beyond masking, absorbing, eliminating, and oxidizing; the most important application is to indoor air quality.  Furthermore, since human food aversion and food intake behavior can have significant consequence for health, well-being, and happiness, ways to steganographically hide one food inside another can be quite powerful.  Although the signal processing results are promising, it remains to validate the efficacy of these methods with experimental tests using human subjects.

\bibliography{abrv,conf_abrv,olfactory_sp}

\begin{thebibliography}{10}
\bibitem{Stinson2014}
L.~Stinson, This phone lets you send smells, not texts, Wired.
\end{thebibliography}

\end{document}